\documentclass[12pt,letterpaper]{article}
\usepackage[top=0.85in,left=0.75in,footskip=0.75in,marginparwidth=2in]{geometry}

\usepackage[utf8]{inputenc}

\usepackage{nameref,hyperref}

\usepackage[right]{lineno}

\usepackage{microtype}
\DisableLigatures[f]{encoding = *, family = * }
\usepackage{ragged2e}
\usepackage{fourier}
\usepackage{caption}
\usepackage{chngcntr}
\usepackage{stfloats}

\usepackage[superscript]{cite}
\raggedright
\usepackage{changepage}

\usepackage[aboveskip=1pt,labelfont=bf,labelsep=period,singlelinecheck=off]{caption}

\makeatletter
\renewcommand{\@biblabel}[1]{\quad#1.}
\makeatother

\usepackage{lastpage,fancyhdr,graphicx}
\usepackage{epstopdf}
\fancyheadoffset[L]{2.25in}
\fancyfootoffset[L]{2.25in}

\usepackage{color}

\definecolor{Gray}{gray}{.25}

\usepackage{graphicx}

\usepackage{sidecap}

\usepackage{wrapfig}
\usepackage[pscoord]{eso-pic}
\usepackage[fulladjust]{marginnote}
\reversemarginpar

\begin{document}
\vspace*{0.35in}

\begin{flushleft}
{\Large
\textbf\newline{Nanodiamonds based optical-fiber quantum probe for magnetic field and biological sensing }
}
\newline
\\

Yaofei Chen\textsuperscript{1,2},
Qianyu Lin\textsuperscript{1,2},
Hongda Cheng\textsuperscript{1,2},
Yingying Ye\textsuperscript{2},
Gui-Shi Liu\textsuperscript{1,2},
Lei Chen\textsuperscript{1,2},
Yunhan Luo\textsuperscript{1,2,3*}
Zhe Chen\textsuperscript{1,2,3}

\bigskip
\bf{1} Guangdong Provincial Key Laboratory of Optical Fiber Sensing and Communications, Jinan University, Guangzhou 510632, China
\\
\bf{2} Department of Optoelectronic Engineering, Jinan University, Guangzhou 510632, China
\\
\bf{3} Key Laboratory of Optoelectronic Information and Sensing Technologies of Guangdong Higher   Educational Institutes, Jinan University, Guangzhou 510632, China
\bigskip
\\
\bf{*} Corresponding author: Yunhan Luo, yunhanluo@163.com

\end{flushleft}

\section*{Abstract}

\marginpar{
\vspace{0.7cm} 
\color{Gray} 
}

\begin{wrapfigure}[18]{R}{85mm}
\includegraphics[width=85mm]{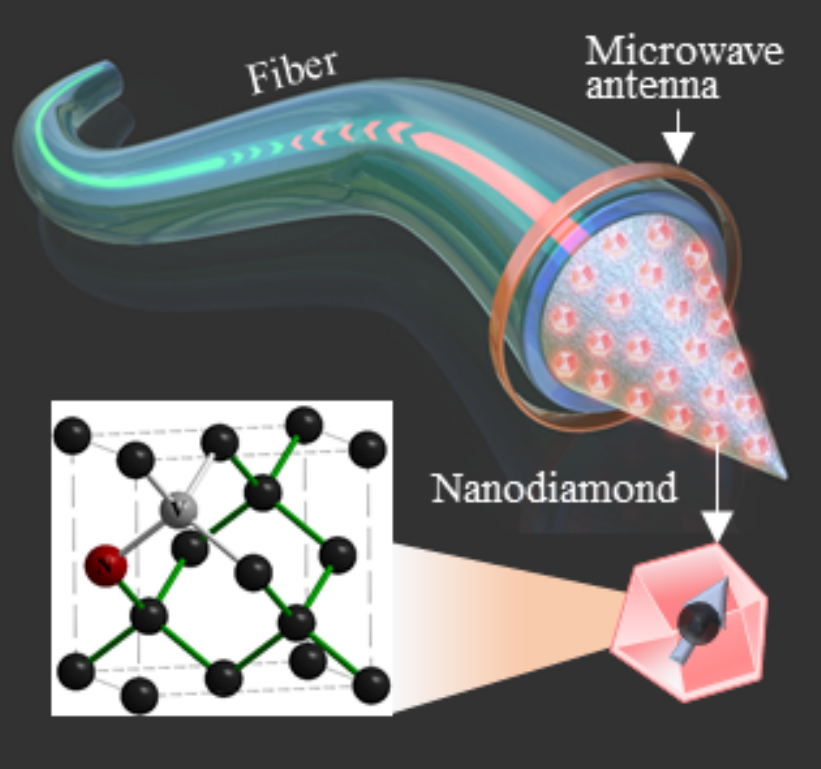}
\captionsetup{labelformat=empty} 
\caption*{} 
\label{fig1} 
\end{wrapfigure}
\justifying
\hspace*{0.4cm}Owing to the unique electronic spin properties, the nitrogen-vacancy (NV) centers hosted in diamond have emerged as a powerful quantum sensor for various physical parameters and biological species. In this work, a miniature optical-fiber quantum probe, configured by chemically-modifying nanodiamonds NV centers on the surface of a cone fiber tip, is developed. Based on continue-wave optically detected magnetic resonance method and lock-in amplifying technique, it is found that the sensing performance of the probe can be engineered by varying the nanodiamonds dispersion concentration and modification duration in the chemical modification process. Combined with a pair of magnetic flux concentrators, the magnetic field detection sensitivity of the probe is significantly enhanced to 0.57 nT/$\mathrm{Hz}^{1 / 2}$ @ 1Hz, a new record among the fiber magnetometers based on nanodiamonds NV. Taking $\mathrm{Gd}^{3+}$ as the demo, the capability of the probe in paramagnetic species detection is also demonstrated experimentally. Our work provides a new approach to develop NV center as quantum probe featuring high integration, miniature size, multifunction, and high sensitivity, etc.
\\KEYWORDS: Quantum probe, cone optical-fiber tip, nanodiamonds, nitrogen-vacancy centers.

\section*{INTRODUCTION}

\hspace*{0.4cm}In recent years, the quantum sensing based on diamond nitrogen-vacancy (NV) centers had gained increasing attentions and developments. The negatively-charged NV (unless specified, NV refers to the negative charge state in this paper) belongs to one of ~500 different color centers discovered in diamond, and it owns the unique properties such as the ideal photostability without photobleaching, optically-readable electronic spin state at room-temperature, and excellent biocompatibility \textsuperscript{\cite{1}}. Most of the parameters that are able to affect the spin state of NV can be easily detected through the optically detected magnetic resonance (ODMR) method. In the aspect of sensing to physical parameters such as magnetic field, temperature, and electric field, the $\mathrm{NV}^{-}$ based sensors have shown great advantages. For example, compared to the well-known superconducting quantum interference device (SQUID) magnetometer, the well-designed NV magnetometers not only possess the comparable sensitivity but can work at room temperature as well \textsuperscript{\cite{2}}. Furthermore, the tetrahedral structure of the diamond crystal lattice enables such magnetometer to directly sense a magnetic field vector, an advantage over the inherently scalar measurements of traditional atomic magnetometers \textsuperscript{\cite{3}}. Employing the nanodiamonds (NDs) containing NV even allows us to measure the intracellular thermal conductivity in situ \textsuperscript{\cite{4}}, which is almost impossible for the traditional sensors. Recently, NV centers had also been proved as a breakthrough tool in the biomedical sensing, though currently it is in its infancy \textsuperscript{\cite{5}}. Utilizing the microwave (MW) or magnetic-field-tunable emission intensity of NV centers and the frequency-domain analysis method, the sensitivity of lateral flow immunoassays succeeded to be improved to an unprecedented level \textsuperscript{\cite{6}}, enabling the single-copy detection of HIV-1 RNA \textsuperscript{\cite{7}}. Particularly, the NV centers have shown great potentials for the detection of the physiologically relevant species (e.g. free radicals, paramagnetic molecules, and ions) \textsuperscript{\cite{8,9,10}} and pH \textsuperscript{\cite{11,12}}, which play an essential role in various critical biological processes, in a label-free manner. For example, Y. Ninio et.al realized a high sensitivity detection to hydroxyl radical by exploiting the fluorescence difference between NV charged states \textsuperscript{\cite{8}}. The detection limit down to the nanomolar range for free radicals had also been achieved while adopting the T1 relaxation measurement \textsuperscript{\cite{9}}. This further provides a promising method to exploring the localized chemical events \textsuperscript{\cite{13}} or chemical redox processes\textsuperscript{\cite{14}}, in which the free radicals or paramagnetic species are incorporated. In a word, the NV centers hosted in diamond have been demonstrated as an ideal quantum probe for the detection to various physical parameters and biological species.
\\
\hspace*{0.4cm}During the implementation of $\mathrm{NV}^{-}$ based sensor, the generally-adopted form is space light coupling \textsuperscript{\cite{15}}, i.e. using a microscope objective or lens to focus the pump laser on the diamond NV, and the emitted fluorescence is collected by the same or additional device. This form is flexible and easy to implement, but the integration of the resulted sensor is poor. Although the recent works had made a big progress on the compactness and portability \textsuperscript{\cite{16,17}}, it is at the expense of spatial resolution, since a long-distance separation between the sensing head and the required optical elements (such as the lens, filter, dichroic mirror) is infeasible for the space light coupling form. A promising solution is to employ the optical-fiber integration form, which exploits the unique features of fiber (e.g. flexible optical path, enabling long-distance light transmission, and small size) and guarantees the high spatial resolutions and integration at the same time. The optical-fiber integration form had been successfully implemented by adhering a diamond block or particle on the fiber facet \textsuperscript{\cite{18,19,20,21,22}}, or filling the micron diamond particles suspension into the air holes of a hollow core fiber \textsuperscript{\cite{23}}, or physically-depositing the NDs on the surface of tapered fiber \textsuperscript{\cite{24,25}}. Nevertheless, the sensitivity of this kind of sensors is not high. Especially for the NDs-integrated fiber magnetometers \textsuperscript{\cite{24,25}}, the sensitivity has not broken below 10 
$\mu$T/$\mathrm{Hz}^{1 / 2}$ so far. Moreover, their applications to the biosensing have not been explored. \\
\hspace*{0.4cm}In this paper, we developed a novel optical-fiber quantum probe incorporated with NDs NV centers, and applied it to the detection of magnetic field and paramagnetic species, respectively. The probe was constructed by modifying the NDs onto the surface of a conical fiber tip through chemical covalent bonding. The influence of two key parameters, i.e. the NDs dispersion concentration (DC) and modification duration (MD) used in modification process, were well studied. Combined with a magnetic flux concentrator, the magnetic field detection sensitivity of the probe was successfully enhanced to sub-nT/$\mathrm{Hz}^{1 / 2}$ at 1 Hz, which is the highest sensitivity among the reported fiber magnetometers based on NDs. The capability of the probe in paramagnetic species sensing was also proved experimentally.

\counterwithout{figure}{section}%

\section*{METHODS AND MATERIALS}
\hspace*{0.4cm}The proposed probe [schematically shown by figure 1(a)] is composed of a multi-mode fiber (MMF) with a conical tip, whose surface is modified with NDs carrying NV centers. NV centers are such a kind of defects in diamond lattice that a carbon atom is substituted by a nitrogen atom and a vacancy appears at its adjacent site [Figure 1(b)]. The fundamental for magnetic field sensing lies on the spin-triplet ground state ${ }^{3} \mathrm{~A}_{2}$ of NV and the Zeeman effect [Figure 1(c)]. In the absence of external magnetic field, the doublet ms=±1 states (denoted by |±1>) are degenerate, and an intrinsic separation of $D=2.87 \mathrm{GHz}$ occurs between the |±1> and singlet |0> resulting from the spin-spin interaction. Once an external magnetic field is applied, the originally degenerate |±1> will split due to the Zeeman effect, and the splitting degree $\Delta v$ is proportional to the magnetic field projected along the NV symmetry axis (denoted by $B_{\mathrm{NV}}$), which can be expressed as $\Delta v=2 \gamma_{\mathrm{e}} B_{\mathrm{NV}}$, where $\gamma_{\mathrm{e}}=28 \mathrm{GHz} / \mathrm{T}$ is the NV gyromagnetic ratio.
\\
\hspace*{0.4cm}The spin-state-dependent fluorescence intensity of NV allows us to exactly measure the $\Delta v$ via the ODMR method, thus obtaining the BNV. To achieve this, a green laser is used to pump the NV from ground state ${ }^{3} \mathrm{A}_{2}$ to the excited state ${ }^{3} \mathrm{E}$ obeying spin conservation, and then decays back the ground state while emitting red fluorescence. Different from that the excited |0> totally returns back the ground |0>, a part of excited |±1> will return to the ground |0> through the non-radiative path, meaning that the |±1> emits a darker fluorescence and NV will be ultimately polarized to the ground |0> by laser pump. At the same time, if a MW is exerted on the NV and its frequency is matched with that between |0> and |+1> (or |-1>), the |0> will be partly depolarized to |+1> (or |-1>), resulting in the decrease of fluorescence. Consequently, by monitoring the fluorescence while sweeping the MW frequency, we can obtain the ODMR spectrum [as shown in Figure 1(d)], which features two dips centered at the frequency $v_{+}$ and $v_{}$, corresponding to the transitions of |0> $\longleftrightarrow$ |+1> and |0> $\longleftrightarrow$ |-1>. Thus, the $\Delta v$=| $v_{+}$ -  $v_{ -}$ | as well as the $B_{\mathrm{NV}}$can be achieved.
\\
\hspace*{0.4cm}The used conical optical-fiber tip, fabricated by chemically-etching a step-index silica MMF (SI2014-N, YOFC, with the core/cladding diameter of 105/125 $\mu \mathrm{m}$), was observed under an optical microscope (MJ30, Guangzhou Mshot Co.), and the taper length was measured as ~270 $\mu \mathrm{m}$ [Figure 1(e)]. The fluorescent carboxylated-NDs (purchased from FND Biotech. Inc.) were characterized by scanning electron microscope (SEM), fourier transform infrared spectroscopy (FTIR, Shimadzu), and photoluminescence spectroscopy (Horiba scientific), respectively. The SEM image [Figure 1(f)] indicates that the NDs have the cube-like morphology and a diameter ~100 nm. The FTIR spectrum [Figure 1(g)] features the dips located at 1635 and 3428 $\mathrm{cm}^{-1}$ (attributed to O-H stretching vibrations) and 1735 $\mathrm{cm}^{-1}$ (induced by C=O bending vibrations), confirming the modification of carboxyl group on the NDs surface \textsuperscript{\cite{26}}. Besides, other chemical bond features,

\begin{figure*}[htbp]
    \centering
    \includegraphics[width=15cm]{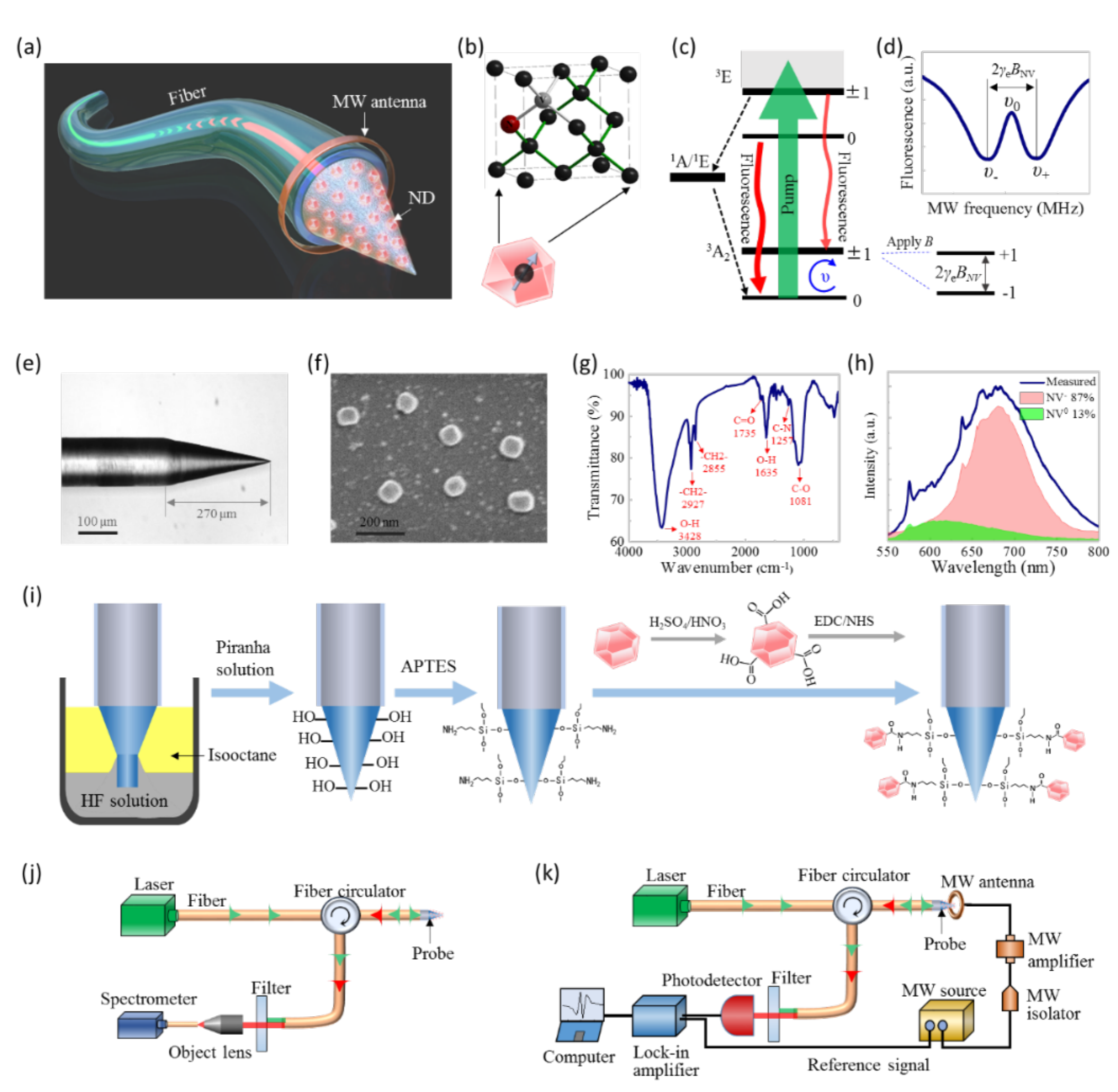}
    \caption{(a) Schematic representation of the optical-fiber quantum probe. It consists of a fiber with conical tip, the NDs modified on the tip, and a MW antenna wound around the tip. (b) The NV defect presented in the diamond lattice. (c) Energy-level structure, including the paths of laser pump and fluorescence emission, MW induced magnetic resonance between |0> and |±1>, and the magnetic field induced split of |±1>, for NV center. (d) Schematic diagram of ODMR spectrum. (e) Microscope image for the bare conical fiber tip. (f)-(h) SEM image, FTIR spectrum, and photoluminescence spectrum (using 532 nm light ford excitation) of NDs. (i) Schematic diagram of probe preparation process. (j) Schematic diagram of the test setup for fluorescence spectrum. (k) Schematic diagram of the measurement setup for lock-in ODMR spectrum. During measurements, the same parameters, including the MW scan range (2820-2920 MHz), scan step (0.5 MHz), dwell time (500 ms), frequency modulation deviation (2 MHz), modulation frequency (500 Hz), lock-in time constant (100 ms), filter roll-off (12 dB/oct), and the gain of the photodetector, etc., were chosen for different probes.}
   \label{fig:1}
\end{figure*}
\newpage
\hspace*{-0.6cm}including the C-O stretching vibrations at 1081 $\mathrm{cm}^{-1}$, C-N bending vibrations at 1257 $\mathrm{cm}^{-1}$ (corresponding to the nitrogen-induced defects in NDs), and the symmetric and antisymmetric stretching vibrations of the CH2 groups at 2855 cm-1 and 2927 cm-1, can also be observed in the FTIR spectrum\textsuperscript{\cite{26}}. The measured photoluminescence spectrum [Figure 1(h)] shows the fluorescence band of NDs ranging from 550 to 800 nm, among which two peaks located at 575 and 637 nm, corresponding to the zero photon lines of NV0 and $\mathrm{NV}^{-}$, are clearly seen. Furthermore, by deconvoluting the measured spectrum into the two pure spectra of NV0 and $\mathrm{NV}^{-}$ \textsuperscript{\cite{27}}, we can evaluate the percentage of $\mathrm{NV}^{-}$ as $87 \%$.
\\
\hspace*{0.4cm}As depicted in Figure 1(i), the probe was fabricated by the following four steps. $I$. A piece of MMF (>40 cm in length) was first cut to obtain a flat end face. Then, the fiber was immersed in the $40 \%$ HF solution covered by a thin layer of isooctane for 1.5 hours. The HF solution reduced the fiber diameter by chemically-etching while the isooctane prevented fiber from being etched, which finally resulted in the formation of a cone fiber tip. The obtained cone tip was sufficiently rinsed by ethanol and deionized water. $II$. Immersed the cone tip in piranha solution (mixture of $\mathrm{H}_{2} \mathrm{SO}_{4}$ and $\mathrm{H}_{2} \mathrm{O}_{2}$ with the volume ratio of 7:3) for 30 minutes to hydroxylate the surface. Then, cleaned the tip using deionized water and dried it under 100 oC. $III$. Incubated the hydroxylated fiber tip in the (3-aminopropyl)-triethoxysilane (APTES) ethanol solution (1:10 in volume) for 5 hours to further aminate the surface, which was followed by sufficiently rinsing using ethanol. $IV$. Modified the NDs on the tip surface. Before modification, The NDs were carboxylated using strong acid (mixture of $\mathrm{H}_{2} \mathrm{SO}_{4}$ and $\mathrm{HNO}_{3}$ with 3:1 in volume) treatment, and the carboxylated NDs were dispersed in water with the aid of sonication treatment at least one hour. To activate the carboxyl groups on the NDs surface, the mixture of N-(-dimethylaminopropyl)-N’-ethylcarbodiimide hydrochloride (EDC, 40 mg/mL) and N-hydroxysuccinimide (NHS, 10 mg/mL) with 1:1 volume ratio was added to the NDs dispersion. After 30 minutes, the aminated cone fiber tip was immersed in the carboxylated NDs dispersion to realize the modification of NDs on the tip surface via amidation reaction. Finally, the modified tip was rinsed using deionized water to remove the unbound NDs, and then dried at $100^{\circ} \mathrm{C}$, thus completing the fabrication of probe. In our experiments, different DCs and MDs were employed to explore the influence of these two parameters on the probe performance. In addition, in order to alleviate the impact of NDs aggregation, the NDs dispersion was sonicated for 10 minutes every 30 minutes during modification.
\\
\hspace*{0.4cm}The employed fluorescence measurement setup for the probes was schematically-presented in Figure 1(j). The 532 nm pump light emitted from a laser (MGL-FN-532nm, CNI Optoelectronics Tech. Co.) was transmitted to the quantum probe through a fiber circulator (Thorlabs WMC3L1S) to excite the fluorescence of NDs. A part of the pump light and the emitted fluorescence from the NDs coupled back the fiber and went through the circulator. A 600 nm long-pass filter was used to isolate the excitation light, and finally, with the aid of an object lens, the remaining fluorescence signal was collected by a spectrometer (AvaSpec-ULS2048XL). For different probes, all test parameters were selected the same to ensure the comparability of test results.
\\
\hspace*{0.4cm}To test the response of the probes to magnetic field via ODMR method, we improved the above setup to that shown in Figure 1(k). The MW, which was supplied from a MW source (Rohde $\&$ Schwarz SMB 100A) and amplified by an amplifier (Mini-Circuits ZHL-16W-43-S+), was applied on the probe through an antenna. The antenna was implemented by winding a copper wire (0.1 mm diameter) around the probe one cycle. Herein, we employed the lock-in amplifying technique to improve the test speed and signal to noise ratio \textsuperscript{\cite{28}}. To implement this, the MW underwent a frequency modulation before emitted from the MW source, and a reference signal synchronized the modulation frequency was sent to the reference channel of a lock-in amplifier (LIA, Sine Scientific Instruments OE1022D), whose signal channel received the fluorescence signal detected by the photodetector (Thorlabs APD410A/M). As a result, by sweeping the microwave frequency and recording the corresponding signal from LIA, we obtained a noise-reduced ODMR spectrum (we called it as LI-ODMR spectrum herein), which was the differential form of the normal spectrum shown in Figure 1(d).

\section*{RESULTS AND DISCUSSIONS}

\subsection*{$Magnetic$ $field$ $sensing$} 
\begin{figure*}[htbp]
    \centering
    \includegraphics[width=15cm]{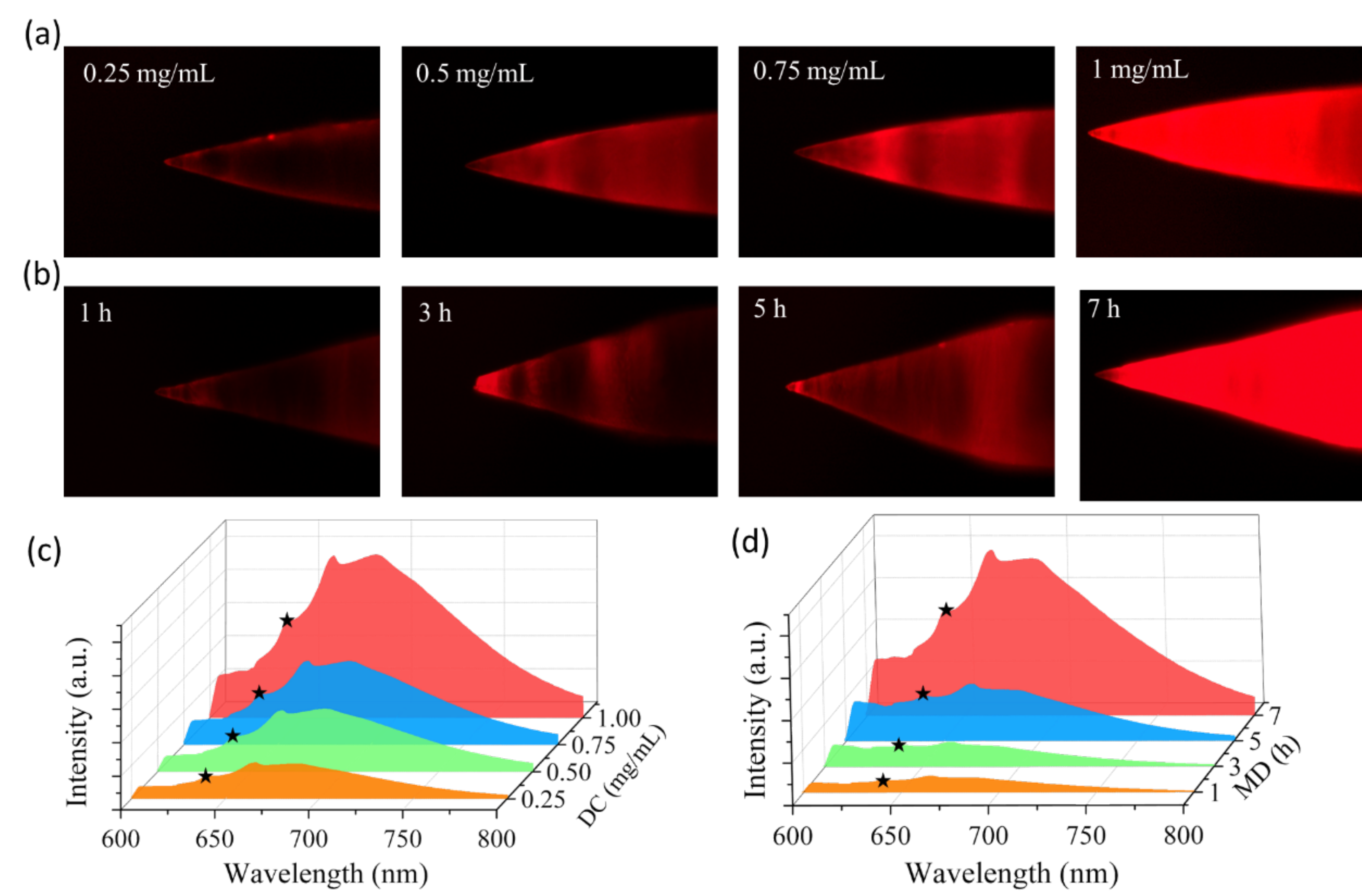}
    \caption{Fluorescence microscope images and corresponding fluorescence spectra of the probes. (a) and (c): the probes modified by different DCs and the same MD of 5 hours. (b) and (d): the probes modified by the same DC of 0.5 mg/mL and different MDs. In (c) and (d), the peaks at 637 nm indicated by the black stars correspond to the zero-phonon line of negatively-charged NV.}
   \label{fig:2}
\end{figure*}

\hspace*{-0.4cm}We first observed the probes, which were fabricated with different DCs and MDs, under a fluorescence microscope (MJ30 customized by adding a fluorescent module, Guangzhou Mshot Co.) using green light excitation and the same parameters. We can observe that the fluorescence emitted by the probe becomes stronger and stronger as the increase of DC and MD [as shown in Figures 2(a)-2(b)], though this part fluorescence fails to make contribution to the useful signal for sensing. To assess the fluorescence collected by the probe itself for sensing, the fluorescence spectra were measured by the setup shown in Figure 1(j). Likewise, the fluorescence that is coupled back fiber and collected by detector improves with the DC and MD, and all the spectra feature an obvious peak at 637 nm, corresponding to the zero-phonon line of negatively-charged NV [Figures 2(c)-2(d)]. All the above phenomena can be attributed to the increase of DC or MD leading to more NDs modified on the conical fiber tip, thus generating a stronger fluorescence.
\\
\hspace*{0.4cm}The LI-ODMR spectra of the probes and their responses to magnetic field variation were tested employing the experimental setup shown in Figure 1(k). Figure 3(a) indicates that as the increase of magnetic field variation, all the spectral lines are centered at ~2.87 GHz while shifting to the both sides, which corresponds to the magnetic-field-induced split between |+1> and |-1> of NV ground stage. The center frequency $v_{0}$, corresponding to the zero-field split between |0> and |±1>, is temperature-dependent with a constant of about -74 kHz/K \textsuperscript{\cite{29}}. During test, $v_{0}$ almost remains unchanged for each probe with a maximal fluctuation of 0.16 MHz, suggesting a temperature variation of 2.2 K [Figure S1(a)]. The response of the resonance frequency $v_{\pm}$ to magnetic field variation, which is calculated by $R v_{\pm}$=abs($\Delta v$/$\Delta B$) for the probes with different DCs are presented in Figures S1(b)-S1(c), and it is indicated that the DC hardly affect the $R v_{\pm}$. This can be attributed to the following two facts: first, the NDs are randomly bond to the probe surface, making the NV axis distributed along every direction in equal probability; second, the split between |+1> and |-1> is essentially determined by the magnetic field projected along the NV axis rather than the number of NV centers.
\\
\hspace*{0.4cm}However, the magnetic-electrical conversion coefficient, which is defined by the $R_{\mathrm{s}}$=max|$\Delta S$($v$) /$\Delta B$|  , i.e. the maximal change slope of the lock-in signal $S(v)$ within the whole magnetic field variation range $\Delta B$ is closely-related to the DC [Figure 2(b)]. The $R_{\mathrm{s}}$ had been enhanced from 2.83 to 15.36 V/T when the DC increased from 0.25 to 1 mg/mL. As shown by Figure 2(c), a higher DC corresponds to a stronger fluorescence, which results in a larger peak-valley contrast [Figure 3(a)], thus an improved $R_{\mathrm{s}}$. To evaluate the magnetic field detection sensitivity of the probes, we first converted the electric noise $S_{\mathrm{N}}(t)$ to magnetic noise $B_{\mathrm{N}}(t)$ in time domain via $B_{\mathrm{N}}(t)$=$S_{\mathrm{N}}(t)$/$R_{\mathrm{s}}$, where the $S_{\mathrm{N}}(t)$ was acquired by consistently sampling the signal (at 1 kHz) from the lock-in amplifier for 60 s at an off-resonance MW frequency. Then, the magnetic noise amplitude spectral density (ASD) was calculated by the Welch’s method \textsuperscript{\cite{2}}. The results [Figure 3(c)] suggest that the detection sensitivity, defined as the magnetic noise spectral density at 1 Hz, improves with the DC, and it reaches 104.97 nT/$\mathrm{Hz}^{1 / 2}$ when the DC increases to 1 mg/mL while MD is kept as 5 hours.
\\
\hspace*{0.4cm}The influence of MD on the probe’s magnetic field responses (including the resonance frequency shift, magnetic-electrical conversion coefficient, and detection sensitivity) shows the same laws as the DC (please see Figure S2 in supporting information for details). Using the 0.5 mg/mL DC, the magnetic-electrical coefficient is improved from 0.88 to 13.22 V/T when the MD rises from 1 to 7 hours [Figure 3(d)]. Correspondingly, the detection sensitivity is enhanced from 3121.79 to 191.22 nT/$\mathrm{Hz}^{1 / 2}$ [Figure 3(e)].
\\
\hspace*{0.4cm}To further enhance the magnetic detection sensitivity, a pair of magnetic flux concentrators (MFCs) were applied to the probes. The MFCs, which are usually made of high permeability materials and specially-designed in the shape, can collect the environmental magnetic flux and focus them to probes by an extremely-enhanced density \textsuperscript{\cite{30}}, thus improving the sensibility to the environmental magnetic field. Herein, we employed two identical cone shape concentrators (base diameter ~30 mm, tip diameter ~0.8 mm, height ~30 mm) made of permalloy (1J79, permeability ~20000), and clamped the probe between the two concentrator tips with a distance ~200 $\mu \mathrm{m}$, as shown by Figure 3(f). For each probe modified by a given DC and MD, its magnetic field response is significantly-enhanced by the MFCs in all aspects, i.e. resonance frequency shift, magnetic-electrical coefficient, and detection sensitivity. Please see the Figures S4-S5 for details. We pick out and present the magnetic response of the probe that is modified by 1 mg/mL DC and 5 hours MD in Figures 3(g)-3(i). The magnetic-electrical coefficient and sensitivity have been improved to 1458.66 V/T and 0.57 nT/$\mathrm{Hz}^{1 / 2}$, the highest ones obtained in our experiments. Figure 3(j) clearly presents the dependence of sensitivity on DC and MD, indicating about two orders of magnitude improvement in sensitivity induced by
\begin{figure*}[htbp]
    \centering
    \includegraphics[width=15cm]{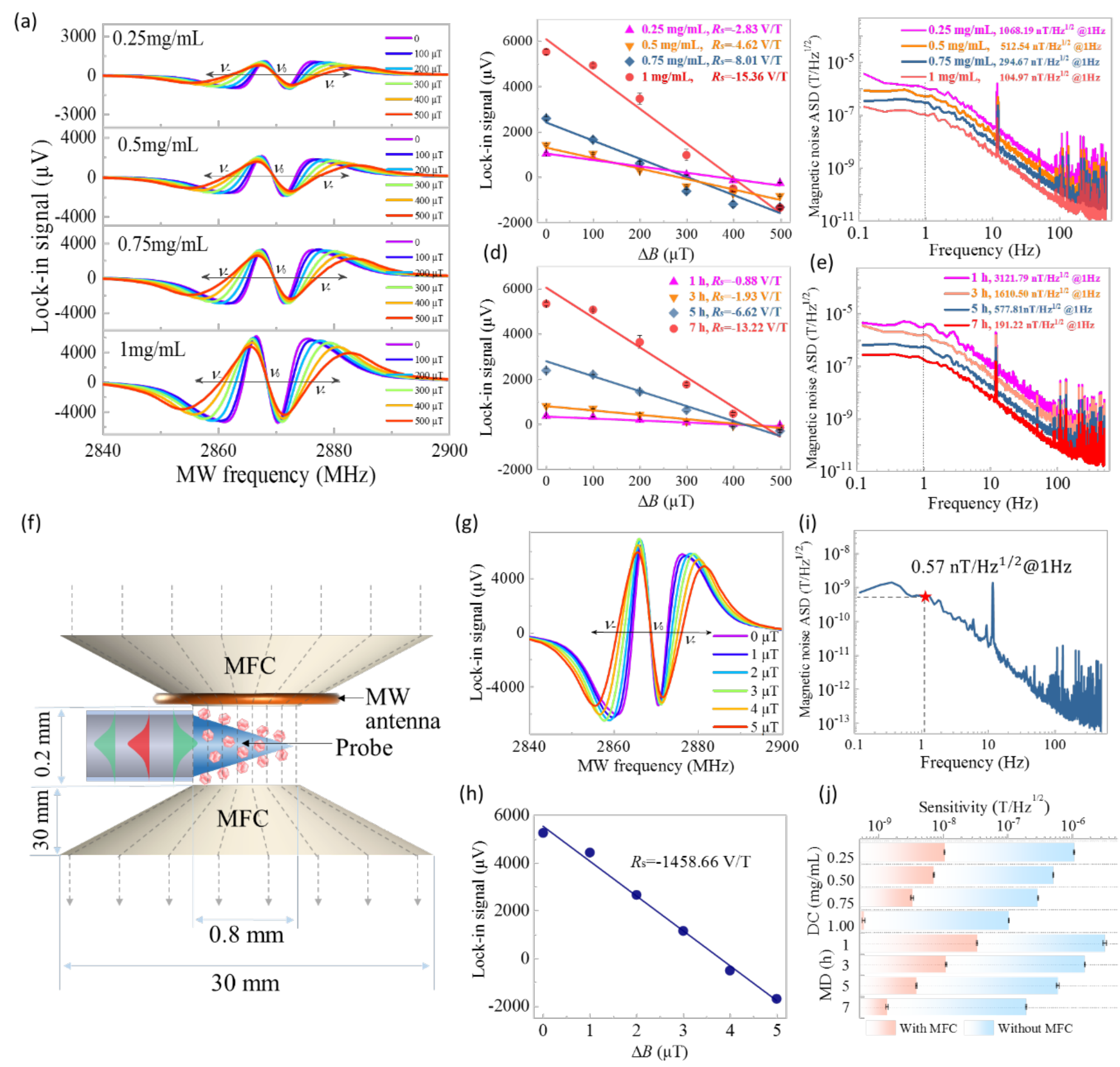}
    \caption{(a) Measured LI-ODMR spectral responses to the magnetic field variation, (b) magnetic-electrical conversion coefficients, and (c) magnetic noise ASD for the probes modified by different DCs and the same MD of 5 hours. (c) The magnetic-electrical conversion coefficients and (d) magnetic noise ASD for the probes modified by the same DC of 0.5 mg/mL and different MDs. (f) Schematic diagram for the application of a pair of MFCs to the probe and the arrangement form. The dash lines represent the magnetic induction lines. (g)-(i) Measured results of the magnetic field response after using MFCs for the 1 mg/mL DC and 7 hours MD probe. Note that the magnetic field variation range is 0-5 $\mu \mathrm{T}$, scaled down by 100 times than the case without using MFCs. (j) Comparison in detection sensitivity with or without the use of MFCs for the various probes.}
   \label{fig:3}
\end{figure*}
\newpage
\hspace*{-0.6cm}MFCs for each probe.
\\
\hspace*{0.4cm}To prove the advantage of using a cone fiber tip, we prepared and measured the sensing performance of a probe with flat end face. At 1 mg/mL DC and 7 hours MD, magnetic-electrical coefficient and detection sensitivity for the flat-tip-based probe are 0.43 V/T and 8596 nT/$\mathrm{Hz}^{1 / 2}$ (Figure S5), which are much lower or poorer that those of the corresponding cone-tip-probe (1458.66 V/T and 0.57 nT/$\mathrm{Hz}^{1 / 2}$) under the same conditions. This mainly arises from that the cone shape tip provides a larger surface to loading a larger amount of NDs, corresponding to a stronger fluorescence collected by fiber, when compared with the flat tip. Moreover, benefiting from the several measures taken in our work (including the special cone shape tip, the optimization on DC and MD, and the application of MFC), the sensitivity of 0.57 nT/$\mathrm{Hz}^{1 / 2}$ obtained herein shows a big improvement than those at $\mu \mathrm{T}$/$\mathrm{Hz}^{1 / 2}$ level obtained in other articles that also employ the fiber integrated with NDs scheme\textsuperscript{\cite{24,25,31,32}} . At the same time, we can look forward to the further enhancement in sensitivity by optimizing the fiber cone parameters (such as the tip length and surface roughness) and the MFCs (in the aspects of materials, structure, size, etc.), which, however, are beyond the scope of this paper and will be presented in our other work.

\subsection*{$Biosensing$ $for$ $paramagnetic$ $species$}
\begin{figure*}[htbp]
    \centering
    \includegraphics[width=15cm]{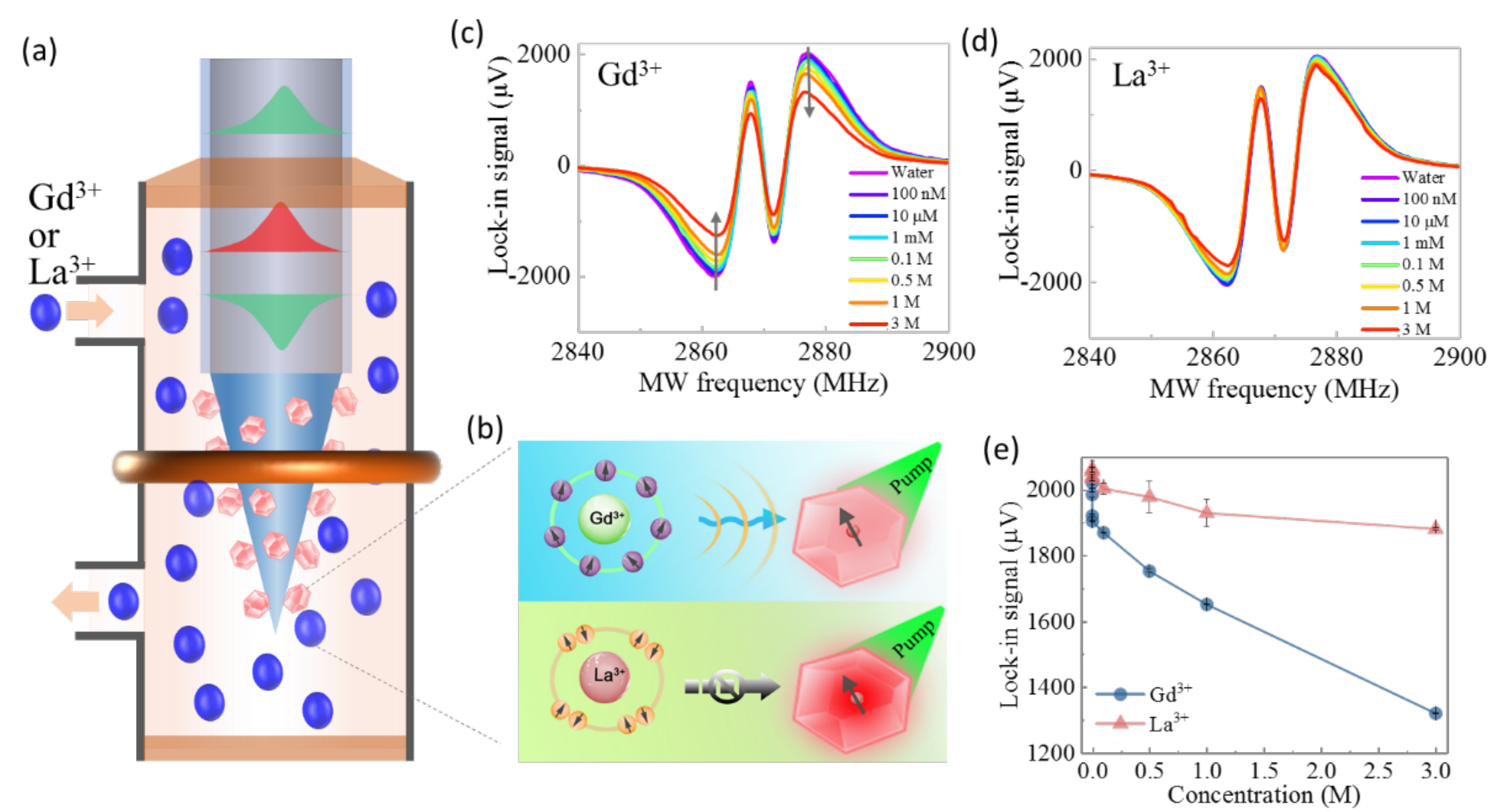}
    \caption{(a) Schematic depiction on the sensing to PS using a microfluidic tube. (b) Schematic illustration for the paramagnetic $\mathrm{Gd}^{3+}$ acting on the NV center via magnetic dipole while the non-paramagnetic $\mathrm{La}^{3+}$  showing no acting. (c)-(d) LI-ODMR spectrum depending on the concentration of $\mathrm{Gd}^{3+}$ and $\mathrm{La}^{3+}$,  respectively. (e) The dependence of lock-in signal on the concentration at the fixed 2877 MHz MW frequency. }
   \label{fig:4}
\end{figure*}

\hspace*{-0.4cm}Paramagnetic species (PS) such as free radicals and paramagnetic metalloproteins,  which contain one or more unpaired electrons in their valence shell, play crucial roles in various physiological processes including tumor growth \textsuperscript{\cite{33}} , immune response \textsuperscript{\cite{34}}, and metabolism \textsuperscript{\cite{35}}, etc. Especially, increasing research suggests that endogenous PS have become the biomarkers for a variety of diseases \textsuperscript{\cite{36,37}}. For example, the paramagnetic metal accumulation in the deep gray matter nuclei was reported to be associated with the Wilson’s disease recently \textsuperscript{\cite{38}}. Reactive oxygen species free radicals have been implicated in an array of diseases including cancer, inflammation, and neurodegenerative diseases such as Parkinson’s disease and Alzheimer’s disease \textsuperscript{\cite{39}}. Therefore, it is of great importance to sense the PS concentration via a suitable detection technology.
\\
\hspace*{0.4cm}Herein, the following experiments were designed to demonstrate the potentials of our probes in detecting PS. As depicted in Figure 4(a), a probe was sealed in a microfluidic tube (with the inner/outer diameter of 2/3 mm) and the copper wire antenna was winded around the tube. Different concentrations of gadolinium nitrate [$\mathrm{Gd}\left(\mathrm{NO}_{3}\right)_{3}$] solutions were successively flowed through the tube, and the corresponding LI-ODMR spectra were recorded employing the same configuration shown in Figure 1(k). The $\mathrm{Gd}^{3+}$, which was widely used as NMR contrast agent, possesses seven unpaired electron in the 4f subshell thus exhibiting strong paramagnetism. Considering that the magnetic spin noise generating from $\mathrm{Gd}^{3+}$ has a broadband spectral density extending to the GHz range, with frequency components corresponding to the NV Larmor precession, an affection from the surrounding $\mathrm{Gd}^{3+}$ will be exerted on the NV, which is consistently polarized to |0> by green laser pump, through the magnetic dipole coupling 10, as schematically-illustrated in Figure 4(b). This consequently reduces the |0> polarizability, manifesting as the decrease of the fluorescence emitted by NV under continue pump. As a control case, the same concentrations of lanthanum chloride ($\mathrm{LaCl}_{3}$) solutions, where the $\mathrm{La}^{3+}$ is chemically similar to $\mathrm{Gd}^{3+}$ but it does not have any unpaired electrons (namely non-paramagnetic), were flowed through as well.
\\
\hspace*{0.4cm}The measured results [Figures 4(c)-4(e)] well support the analysis as illustrated above. The contrast of the LI-OMDR spectrum, embodiment of the fluorescence intensity generated from NV under continue pump, exhibits significant dependence on the $\mathrm{Gd}^{3+}$ concentration while suffering little effect from $\mathrm{La}^{3+}$. Specifically, as the increase of $\mathrm{Gd}^{3+}$ concentration, the contrast reduces obviously, but the resonant frequencies remain unchanged. The latter is attributed to the location of $\mathrm{Gd}^{3+}$ magnetic spin noise in the high frequency region (GHz), which affects the NV in a way of polarization reduction, rather than the detectable split of ground |±1> induced by a static or slowly-varying magnetic field using the current continue-wave ODMR method. We also note a $\mathrm{La}^{3+}$-related slight contrast variation, which is much smaller compared to the $\mathrm{Gd}^{3+}$. This suggests that in our case of continue-wave ODMR, the magnetic spin noise (from the paramagnetic $\mathrm{Gd}^{3+}$) plays a dominant role in the contrast reduction, whereas the another well-known process of charge state dynamic \textsuperscript{\cite{40}} that contributes to the $\mathrm{La}^{3+}$-related slight contrast variation is suppressed. Based on the similar evaluation method for sensitivity as the magnetic field, the detection sensitivity for $\mathrm{Gd}^{3+}$ is evaluated as 26.8 nM/$\mathrm{Hz}^{1 / 2}$ (Figure S6).
\\
\hspace*{0.4cm}In a word, we demonstrated a novel scheme for PS detection using an optical-fiber quantum probe featuring high integration, small size, and high sensitivity. Moreover, the employed lock-in technique and ODMR method enable us to acquire the results with high signal-to-noise ratio in a short time (hundreds of milliseconds), which shows a big advantage when compared with the longitudinal relaxation ($T$1) measurement method (generally needs a few minutes even more) \textsuperscript{\cite{9,35}}. Therefore, the method presented here holds great potentials in studying the PS-related physiological or chemical dynamic processes within a local region. It should be pointed out that the current work focuses to demonstrate the capability of the proposed probe in PS sensing, the further work aimed at specificity and mechanism clarity is necessary before practical application.

\section*{CONCLUSIONS}

\hspace*{0.4cm}In summary, a novel configuration to exploiting NV centers in diamond as a quantum probe is proposed and demonstrated. The probe is implemented by anchoring the nanodiamonds containing NV centers onto a cone fiber tip via chemical covalent bonding, and it is applied to the sensing to magnetic field and PS based on the continue-wave ODMR method and lock-in amplifying technique. Through experimental measurements, it is found that both of the DC and MD have little influence on the responses of center and resonant frequencies to the magnetic field variation. Whereas, the magnetic-electrical conversion coefficient and detection sensitivity are closely-related to the DC and MD. Increasing DC and MD is beneficial to the improvement of magnetic detection sensitivity, because more NDs can be modified on the fiber tip thus enhancing the fluorescence signal. Additionally, after applying a pair of MFCs to the probe, the sensitivity succeeds to be further enhanced with two orders of magnitude. The highest sensitivity of 0.57 nT/$\mathrm{Hz}^{1 / 2}$ @ 1Hz had been achieved in our experiments, and we can look forward to a further enhancement by optimizing the fiber cone parameters (tip length, surface roughness, etc.) and the MFCs (in the aspects of materials, structure, size, etc.). The probe also exhibits great potentials in biosensing to PS, which was demonstrated by a well-designed experiment where the probe was much sensitive to the paramagnetic $\mathrm{Gd}^{3+}$ but little affected by the non-paramagnetic $\mathrm{La}^{3+}$. The detection sensitivity for $\mathrm{Gd}^{3+}$ was achieved as 26.8 nM/$\mathrm{Hz}^{1 / 2}$ in experiments. Our work paves the way to the development of optical-fiber based quantum probe featuring high integration, miniature size, multifunction, and high sensitivity.

\newpage

\section*{Supporting Information}
\begin{figure*}[htbp]
    \centering
    \includegraphics[width=15cm]{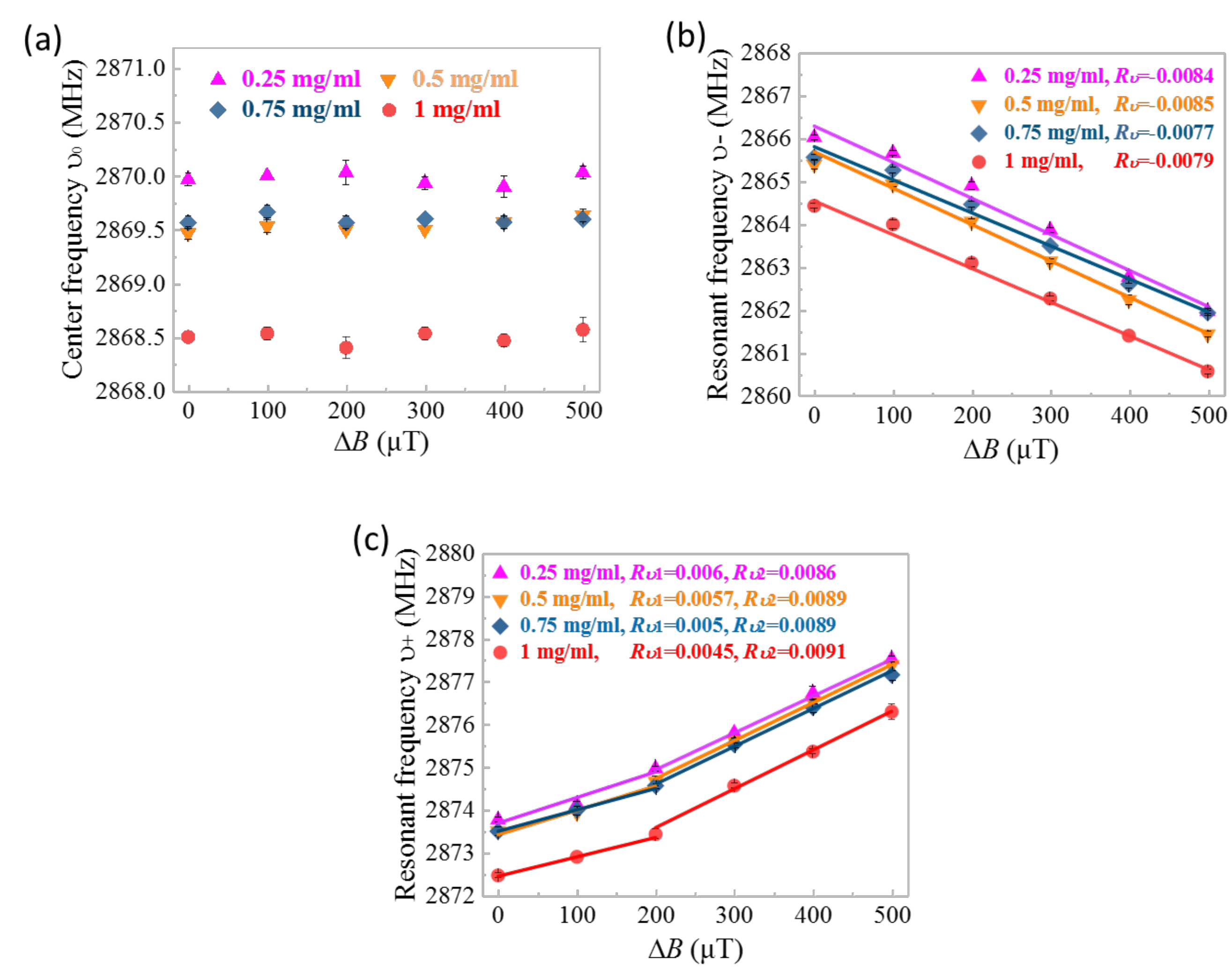}
    
   \label{fig:S1}
\end{figure*}

\hspace*{-0.6cm}{\bf Figure S1.}\textbf{}Center and resonant frequencies dependence on the magnetic field variation for the probes with different DCs and the same MD of 5 hours. (a) Center frequency $\mathbf{v}_{0}$. The maximal deviation of 0.16 MHz occurs between the $\Delta B$at 200 and 500 $\mu \mathrm{T}$for the probe with DC 1 mg/mL, corresponding to a temperature variation of 2.2 K according the temperature-dependent constant of about -74 kHz/K \textsuperscript{\cite{29}}. The $\mathbf{v}_{0}$ deviation among the probes, the maximum of ~1.5 MHz occurring between the probes with DC 0.25 and 1 mg/mL, is mainly due to the temperature variation and measurement error during experiments, and the variations in static local strain and transverse magnetic field in NDs for the different probes \textsuperscript{\cite{41}}. (b) Resonant frequency $R_{v-}$. The linear fitting results show that the $R_{v-}$=abs( $\Delta v_{-}$/ $\Delta B$) are 0.0084, 0.0085, 0.0077, and 0.0079 MHz/$\mu \mathrm{T}$, with a maximal deviation of $8.3 \%$, for the probes with different DCs. This suggests that the DC has little influence on the response of $v-$ to magnetic field, namely $R_{v-}$. (c) Resonant frequency $v_{+}$. The linear fitting results show that the $R_{v+}$=abs( $\Delta v_{+}$/ $\Delta B$) are almost equal for the probes with different DCs. This suggests that the DC also has little influence on the response of $v_{+}$ to magnetic field, namely $R_{v+}$.
\newpage
\begin{figure*}[htbp]
    \centering
    \includegraphics[width=15cm]{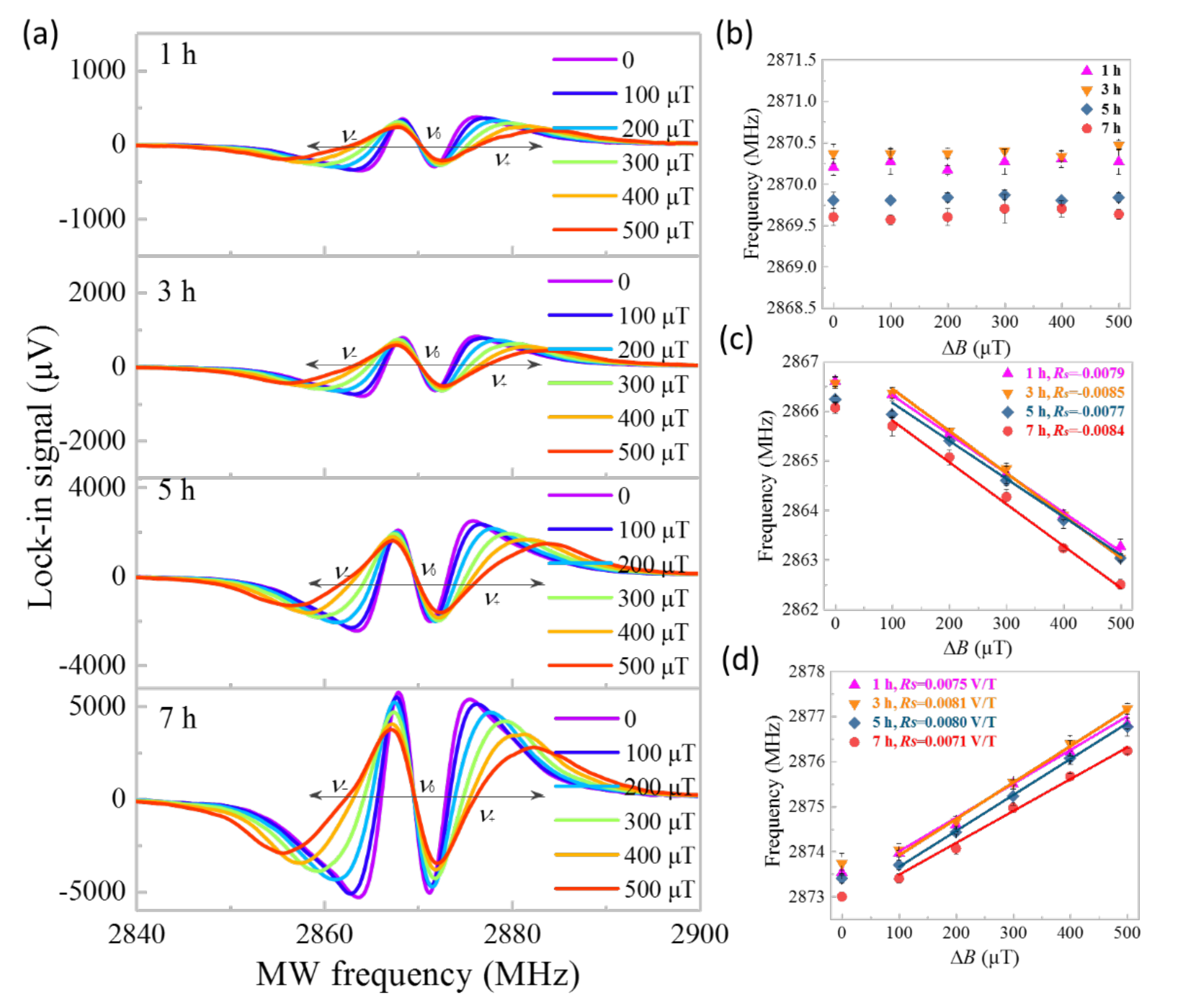}
    
   \label{fig:S2}
\end{figure*}

\hspace*{-0.6cm}{\bf Figure S2.}\textbf{} (a) Measured LI-ODMR spectral responses to the magnetic field variation for the probes modified by the same DC of 0.5 mg/mL and different MDs. Similarly, as the increase of magnetic field, all the spectral lines are centered at ~2.87GHz while shifting to the opposite sides. Meanwhile, the peak-valley contrast increases as well, which corresponds to a stronger fluorescence at a larger MD as shown by Figure 2(c). Therefore, both the magnetic-electrical conversion coefficient and detection sensitivity improve with the MD, as shown in Figure 3(d)-3(e). (b) Center frequency $\mathbf{v}_{0}$, (c) resonant frequency $R_{v-}$, and (c) resonant frequency $v_{+}$ dependence on the magnetic field variation for the different probes. The similar conclusions as illustrated by the above Figure S1 can be obtained: the temperature-dependent center frequency $\mathbf{v}_{0}$ remains almost unchanged during the test for each probe; the MD has little influence on the response of the resonant frequencies $v_{+}$ and $R_{v-}$ to magnetic field variation.

\begin{figure*}[htbp]
    \centering
    \includegraphics[width=15cm]{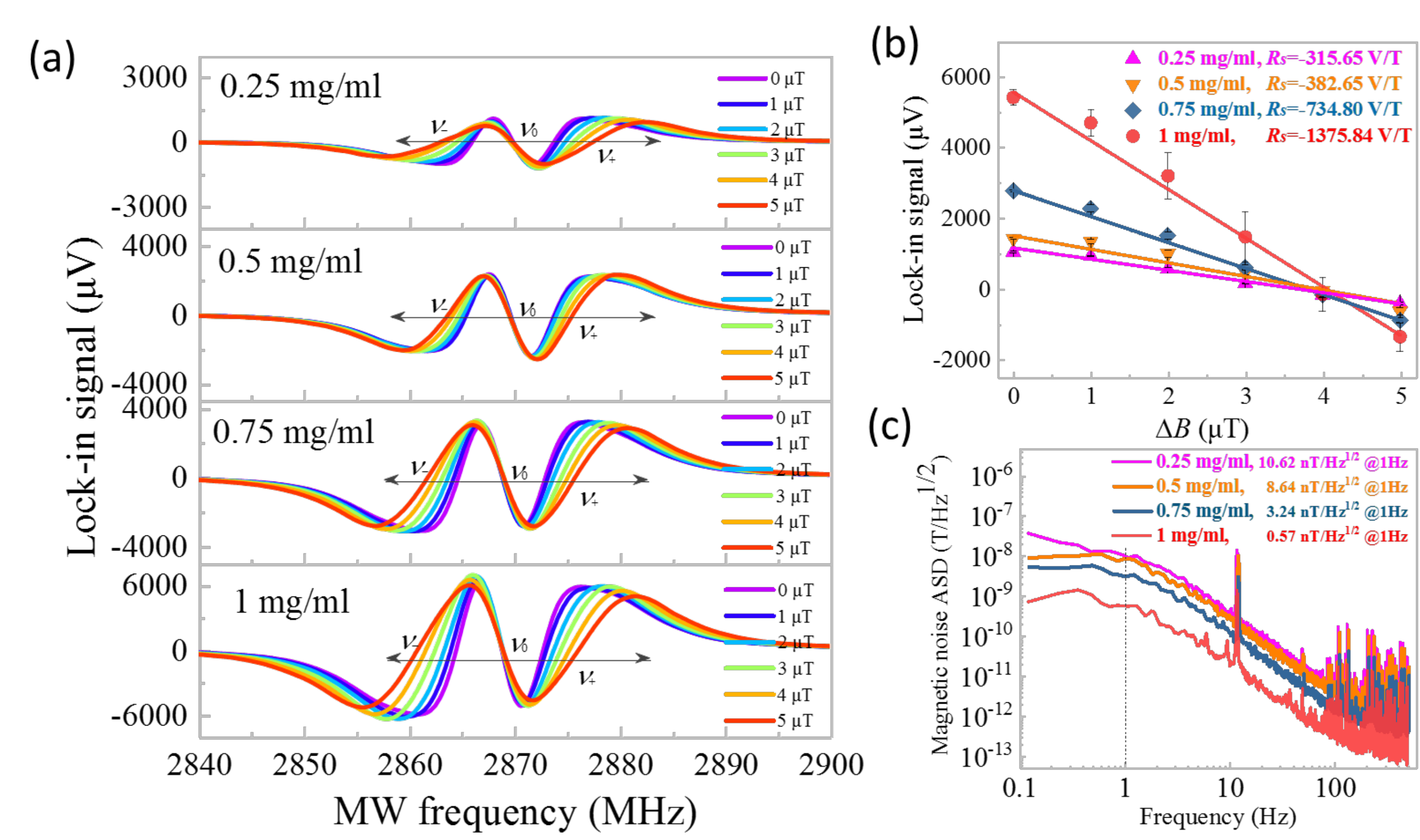}
    
   \label{fig:S3}
\end{figure*}
\newpage
\hspace*{-0.6cm}{\bf Figure S3.}\textbf{} After using MFCs, the measured responses to magnetic field variation for the probes with the different DCs and the same MD of 5 hours. (a) LI-ODMR spectral responses to the magnetic field variation. Note that though the magnetic field variation range is scaled down 100 times (i.e. 0-5 $\mu \mathrm{T}$) when compared to case without using MFCs (i.e. 0-500 $\mu \mathrm{T}$), the significant spectral response can be observed. (b)-(c) Magnetic-electrical conversion coefficients and magnetic noise ASD for each probe. These two performance indicators both have improvement by two orders of magnitude after using MFCs, compared with the results in Figures 3(b)-3(c).
\newpage
\begin{figure*}[htbp]
    \centering
    \includegraphics[width=15cm]{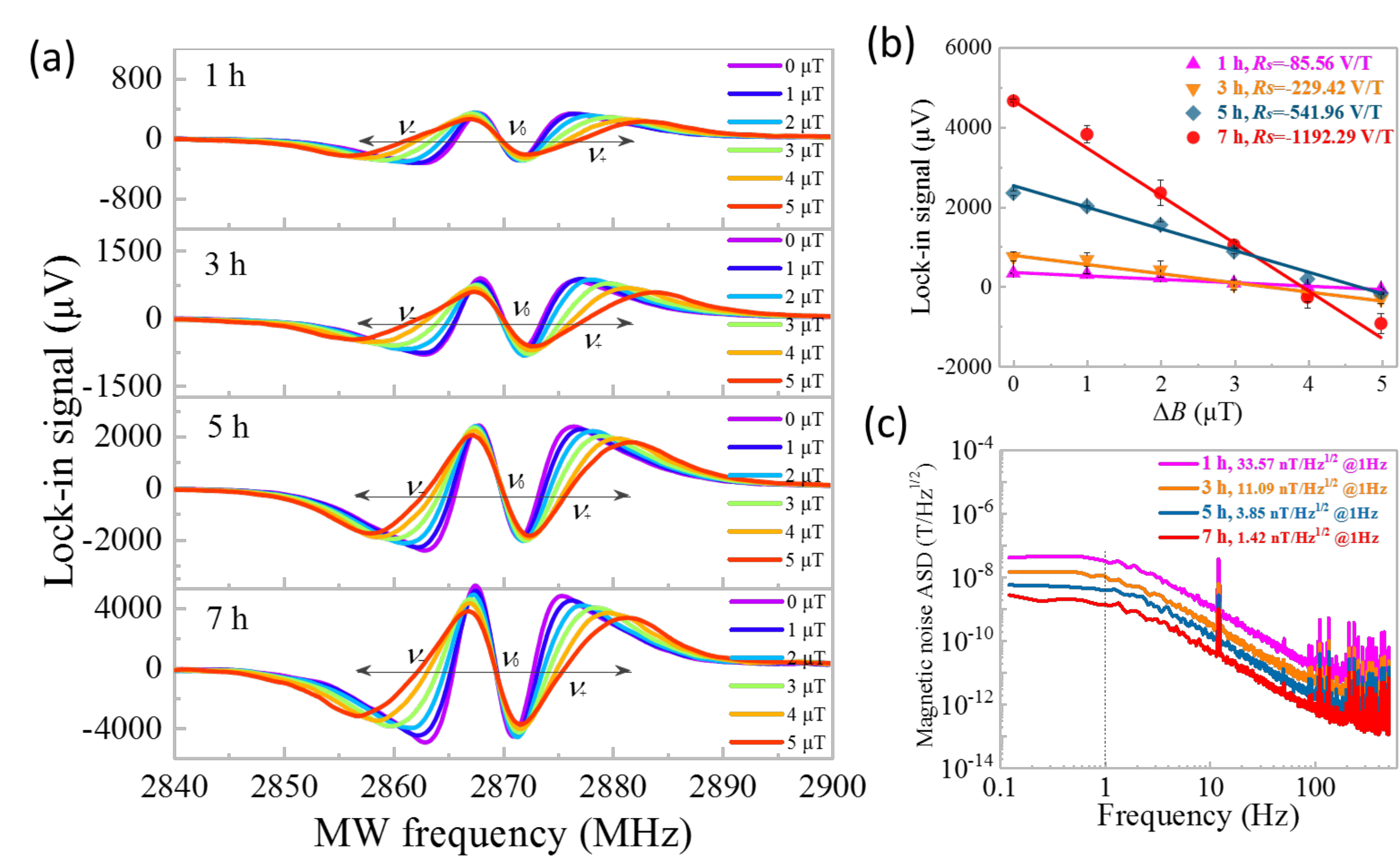}
    
   \label{fig:S4}
\end{figure*}
\hspace*{-0.6cm}{\bf Figure S4.}\textbf{} After using MFCs, the measured responses to magnetic field variation for the probes with the same DC of 0.5 mg/mL and the different MDs. (a) LI-ODMR spectral responses to the magnetic field variation. Note that though the magnetic field variation range is scaled down 100 times (i.e. 0-5 $\mu \mathrm{T}$) when compared to case without using MFCs (i.e. 0-500 $\mu \mathrm{T}$), the significant spectral response can be observed. (b)-(c) Magnetic-electrical conversion coefficients and magnetic noise ASD for each probe. These two performance indicators both have improvement by two orders of magnitude after using MFCs, compared with the results in Figures 3(d)-3(e).

\newpage
\begin{figure*}[htbp]
    \centering
    \includegraphics[width=15cm]{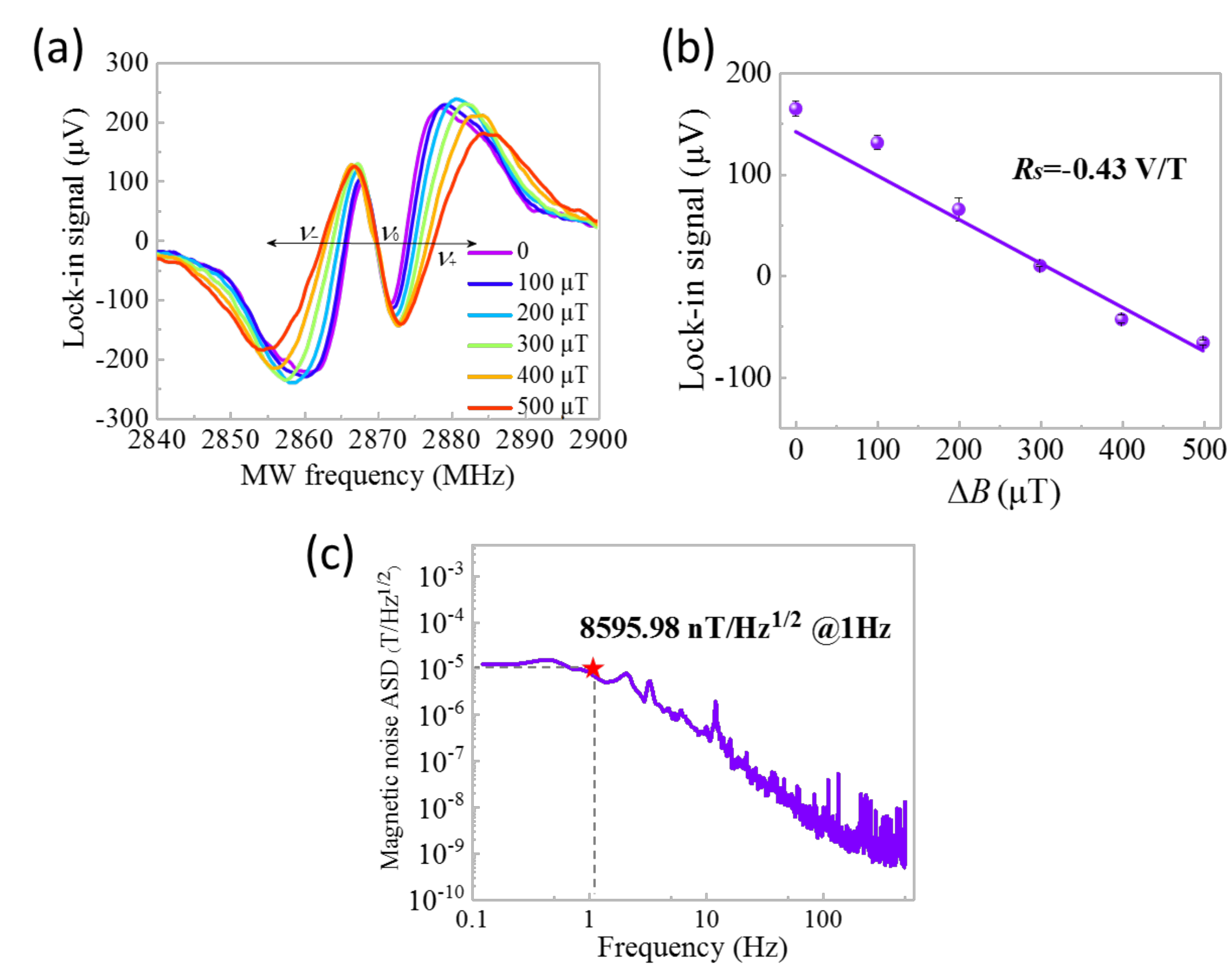}
    
   \label{fig:S5}
\end{figure*}
\hspace*{-0.6cm}{\bf Figure S5.}\textbf{} Measured magnetic field sensing results for the fiber probe with a flat end face modified by 1 mg/mL DC and 7 hours MD. (a) LI-ODMR spectral responses to the magnetic field variation. (b) Measured magnetic-electrical conversion coefficient and (c) magnetic noise ASD for the probe. Note that though the LI-ODMR spectrum exhibits the same magnetic-response behaviors, i.e. centered at ~2.87GHz while shifting to the opposite sides as the increase of magnetic field variation, the peak-valley contrast is significantly-reduced when compared with the cone-tip probe fabricated at the same conditions. Correspondingly, both the magnetic-electrical conversion coefficient and the detection sensitivity have deteriorated.

\newpage
\begin{figure*}[htbp]
    \centering
    \includegraphics[width=15cm]{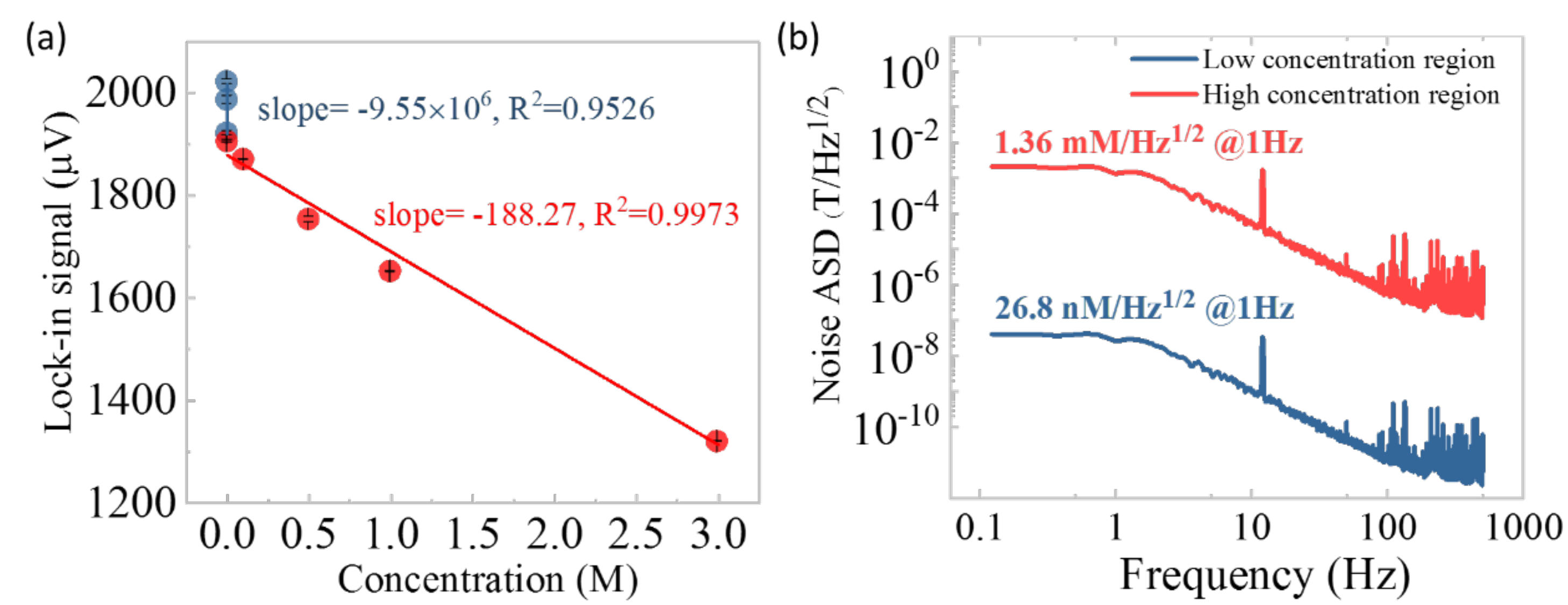}
    
   \label{fig:S6}
\end{figure*}
\hspace*{-0.6cm}{\bf Figure S6.}\textbf{} (a) Linear fittings of the measured data for $\mathrm{Gd}^{3+}$ at the linear regions. The obtained responsivity are $-9.55 \times 10^{6}$ $\mu \mathrm{V}$/M and -188.27 $\mu \mathrm{V}$/M at the low and high concentration regions. (b) Noise ASD for the $\mathrm{Gd}^{3+}$ detection. The noise ASD herein is obtained through Welch’s method same as the evaluation for magnetic noise ASD presented in text. Results indicate that the probe has the detection sensitivities of 26.8 nM/$\mathrm{Hz}^{1 / 2}$ and 1.36 mM/$\mathrm{Hz}^{1 / 2}$ at the low (0-10 $\mu \mathrm{M}$) and high (1mM-3M) concentration regions, respectively.

\section*{ASSOCIATED CONTENT}
\vspace{+0.4cm}
\section*{ORCID}
Yaofei Chen: 0000-0002-7067-6562
\\
Yunhan Luo: 0000-0002-6587-832X
\section*{Notes}
The author declare no competing financial interest.

\section*{Acknowledgments}
This work is supported by the National Natural Science Foundation of China (NSFC) (61805108, 62175094, 61904067, 62075088); Guangdong Basic and Applied Basic Research Foundation (2020A1515011498, \\2017A010101013); Science $\&$ Technology Project of Guangzhou (201707010500,201807010077, 201704030105, 201605030002); Science and Technology R$\&$D Project of Shenzhen (JSGG20201102163800003, \\JSGG20210713091806021); Fundamental Research Funds for the Central Universities (21620328).
\nolinenumbers

\newpage
\bibliography{library}

\bibliographystyle{unsrt}

\end{document}